# Realizing Ultra-Fast and Energy-Efficient Baseband Processing Using Analogue Resistive Switching Memory

Qunsong Zeng*, Jiawei Liu*‡, Jun Lan†, Yi Gong‡, Zhongrui Wang*, Yida Li†, and Kaibin Huang*

* Department of Electrical and Electronic Engineering, The University of Hong Kong, Hong Kong S.A.R.

‡ Department of Electrical and Electronic Engineering, Southern University of Science and Technology, Shenzhen, China

† School of Microelectronics, Southern University of Science and Technology, Shenzhen, China

To support emerging applications ranging from holographic communications to extended reality, next-generation mobile wireless communication systems require ultra-fast and energy-efficient (UFEE) baseband processors. Traditional complementary metal-oxide-semiconductor (CMOS)-based baseband processors face two challenges in transistor scaling and the von Neumann bottleneck. To address these challenges, in-memory computing-based baseband processors using resistive random-access memory (RRAM) present an attractive solution. In this paper, we propose and demonstrate RRAM-based in-memory baseband processing for the widely adopted multiple-input-multiple-output orthogonal frequency division multiplexing (MIMO-OFDM) air interface. Its key feature is to execute the key operations, including discrete Fourier transform (DFT) and MIMO detection using linear minimum mean square error (L-MMSE) and zero forcing (ZF), in one-step. In addition, RRAM-based channel estimation as well as mapper/demapper modules are proposed. By prototyping and simulations, we demonstrate that the RRAM-based full-fledged communication system can significantly outperform its CMOS-based counterpart in terms of speed and energy efficiency by $10^3$ and $10^6$ times, respectively. The results pave a potential pathway for RRAM-based in-memory computing to be implemented in the era of the sixth generation (6G) mobile communications.

**Introduction**

While the fifth generation (5G) mobile networks are being deployed, sixth generation (6G) is under development all over the world to provide a new infrastructure for propelling the digital economy forward and realizing Society 5.0[1]. The performance of 6G will be unprecedented as reflected in a set of target key performance indicators (KPIs), dictating a peak data rate to go beyond 100Gb/s, having a minimum latency 0.1ms, and achieving an energy efficiency of $10^{-12}$J/bit[2-6]. This coined the term *ultra-fast-and-energy efficient* (UFEE) communication and will enable a wide range of emerging applications, for example, industrial automation[7,8], tactile internet[9-11], holographic communications[12,13], and digital twin[14,15] (see Fig. 1a). Hence, this provides a strong motivation for 6G researchers to explore the largely unoccupied Terahertz (THz) spectrum[2-6]. However, the required scaling up of sampling rates to the hundreds of GHz level will dramatically increase the power consumption and complexity of baseband processing, making it challenging to realize the 6G vision[16-18]. This is further exacerbated by the increasingly sophisticated communication techniques required, including large-scale *multiple-input-multiple-output* (MIMO), high-dimensional *orthogonal frequency division multiplexing* (OFDM), and interference management. From 2G to 5G era, baseband processing demands have been satisfied largely by shrinking transistor size as governed by Moore's law. Accordingly, the semiconductor industry has evolved from planar bulk Metal-Oxide-Semiconductor Field-Effect Transistors (MOSFETs) to the recent 3D FinFETs and Gate All Around (GAA) architectures to improve transistor performance and density in an integrated circuit (IC) chip[19]. However, this approach is facing increasing challenges as transistor size approaches the atomic limit[20]. Baseband processing and computing at large face two bottlenecks: the von Neumann bottleneck and the power wall, incurring large energy and footprint overheads. The former is due to data shuttling between the physically separated processing and storage units, resulting in significant latency and high energy consumption (e.g., 100-time more than digital logical circuits)[21]. In the latter, the increasing power density of transistors as the transistor size shrinks has created a "power wall" that limits practical processor frequency to ~4 GHz since 2006[22], falling far short of the requirements for THz communications.

In the past decade, researchers have started to improve computing latency and energy consumption by employing an architecture that co-locates data processing and storage, so called *in-memory computing*. Emerging non-volatile memories such as resistive random-access memory (RRAM) is touted as one of the most potential candidates for such computational memory devices[23]. It has been reported that parallel execution of a larger number (e.g., millions) of multiply-and-accumulate (MAC) operations for matrix vector multiplications (MVM) can be accomplished with extremely high energy-efficiency and low latency[21,24]. This makes in-memory computing a UFEE solution for MVM intensive applications such as deep neural networks[25-42], and linear algebra computation (e.g., linear system solver and eigen solver)[43-45]. Such advantages can naturally contribute to the trend of seamless integration of communication and artificial intelligence (AI) for the next-generation Internet-of-Things (IoTs). A new paradigm for communications called *in-memory baseband processing*, which adopts the emerging in-memory computing architecture and novel signal processing approach are potential key factors to alleviate the challenges faced by researchers in realizing UFEE connectivity in the era of 6G.

6G will feature scaling up of different physical-layer technologies, for example, ultra-massive MIMO using antenna arrays with hundreds to thousands of elements, OFDM comprising thousands of sub-carriers, and high-order modulation constellations with thousands of points. The resultant baseband processing will involve frequent large-scale matrix operations. This motivates us to propose the new paradigm of in-memory baseband processing, which relocates the conventional digital operations to the analogue domain to achieve UFEE processing. In this article, we present the design of an in-memory baseband processor for MIMO-OFDM which is a dominant air-interface technology for 5G-and-beyond[2,4,5]. The key novelty includes modules, namely OFDM demodulation, MIMO detection, and channel estimation, implemented using reliable $HfO_2$ RRAM-based in memory computing approach[46-48]. The OFDM module implements the discrete-Fourier transform (DFT) using a RRAM crossbar array. Using such an array to store DFT matrix enables one-step DFT operation, cutting down the power/latency overheads in conventional CMOS-based processor significantly. Furthermore, the required channel matrix inversion for MIMO detection is realized using a novel RRAM circuit featuring stability and easy mode switching. The performance of the design is evaluated using both proof-of-concept prototypes and system simulation. We show that the speed and energy-efficiency can be boosted up to $10^3$ and $10^6$ times respectively as compared to the conventional CMOS-based baseband processors.

**Results:**

**Overview of RRAM-based Baseband Processor**

The proposed system-on-chip (SoC) at intelligent edge clients, equipped with central processing unit (CPU), graphics processing unit (GPU), neural processing unit (NPU), sensor, baseband processor, and radio modems are illustrated in Fig. 1b. Different from traditional CMOS-based processors, the proposed circuit features a hybrid analogue-digital architecture where in-memory computing within analogue RRAM arrays is used to accelerate both baseband processing and AI tasks. The baseband processor considered here targets the MIMO-OFDM air interface, where a pair of multi-antenna transmitter and receiver communicate over a broadband channel. In broadband communications, frequency selective fading occurs when the channel having a coherence bandwidth is smaller than that of the signal causes its distortion. As a popular technology for coping with such fading as well as inter-symbol interference, OFDM is adopted to divide the whole bandwidth into $N_c$ orthogonal sub-channels. As a result, each sub-channel, say the $k$-th sub-channel, is a narrowband channel with $N_t$ transmit and $N_r$ receive antennas, modelled by a MIMO-channel matrix $\mathbf{H}^{(k)} \in \mathbb{C}^{N_r \times N_t}$ that is fixed within an OFDM symbol. The input-output relation of a MIMO system over the $k$-th sub-channel is given as

$$\mathbf{y}^{(k)} = \mathbf{H}^{(k)}\mathbf{x}^{(k)} + \mathbf{z}^{(k)}, \tag{1}$$

where $\mathbf{x}^{(k)} \in \mathbb{C}^{N_t \times 1}$ consists of symbols at the $k$-th sub-carrier, $\mathbf{y}^{(k)} \in \mathbb{C}^{N_r \times 1}$ comprises the received symbols at the $k$-th sub-carrier, and $\mathbf{z}^{(k)}$ represents the additive white Gaussian noise (AWGN) in propagation.

The architectures of the RRAM-based transceiver are illustrated in Fig. 1c and Fig. 1d. The baseband (information) processing starts at the mapper module in the transmitter that transforms bits into symbols and ends at the demapper module in the receiver that transforms the symbols back to bits. Though both the mapper and demapper modules are traditionally implemented by simple logical circuits, we propose the use of RRAM arrays in this regard towards fully analogue bits-to-symbol encoding or decoding (see Supplementary Note 4). The digital modulation is chosen as 16 quadrature amplitude modulation (16-QAM) unless specified otherwise, which maps a 4-bit string to one of the 16 points on the constellation diagram as shown in Fig. 1j. The bit stream is split into in-phase (denoted by $I$) and quadrature (denoted by $Q$) streams, associated with 0-degree and 90-degree phase shifts of the carrier wave, respectively. $I$ and $Q$ components are Gray encoded, i.e., neighbour points only differ in a single bit, to produce symbol points in the constellation. The system performance is evaluated by two metrics: 1) the Modulation Error Ratio (MER) which measures the dispersion of the constellation of the received symbols; 2) the Bit Error Ratio (BER) which is the number of bit errors divided by the total number of transmitted bits. The definitions of the MER and BER are given in Supplementary Note 2. In this work, we focus on the baseband processing between the mapper and demapper, as schematically illustrated by the highlighted modules summarized in Fig. 1c. The module in the transmitter performs inverse DFT (IDFT). For the receiver (see Fig. 1d), the three modules are DFT module, channel estimator, and MIMO detector. To reconcile signals and channels in the complex domain and the fact that RRAM devices store and compute real numbers, we propose to apply the mapping $\mathcal{R}: \mathbb{C}^{K \times L} \to \mathbb{R}^{2K \times 2L}$ which transforms a complex matrix $\mathbf{A} \in \mathbb{C}^{K \times L}$ into a real matrix $\mathcal{R}(\mathbf{A}) = \begin{bmatrix} \Re(\mathbf{A}) & -\Im(\mathbf{A}) \\ \Im(\mathbf{A}) & \Re(\mathbf{A}) \end{bmatrix} \in \mathbb{R}^{2K \times 2L}$. The complex vector is translated as the input voltages (or currents) for the RRAM array, with the mapping $\mathcal{T}: \mathbb{C}^{K \times 1} \to \mathbb{R}^{2K \times 1}$ transforming a complex vector $\mathbf{x} \in \mathbb{C}^{K \times 1}$ into a real vector $\mathcal{T}(\mathbf{x}) = \begin{pmatrix} \Re(\mathbf{x}) \\ \Im(\mathbf{x}) \end{pmatrix} \in \mathbb{R}^{2K \times 1}$. Such transformations are proved to enable the equivalent computation involving complex matrices and vectors (see Supplementary Note 3).

We have used the $HfO_2$-based RRAM arrays as the hardware accelerators for its compatibility with traditional CMOS process and reliable electrical characteristics. Details of the RRAM array fabrication are described in Methods. Fig. 1f shows the fabricated RRAM array, while the zoomed-in Transmission Electron Microscopy (TEM) image of the RRAM device is shown in Fig. 1g. Our fabricated RRAM devices shows uniform DC resistive switching, with a set/reset voltage around 1V under a compliance current of 1mA (Fig. 1h). As a non-volatile analogue device, our RRAM device exhibits multi-level resistance states and excellent data retention to ensure accurate in-memory computing (see Fig. 1i). For matrix mapping, the conductance programming of the HfO$_2$ RRAM device can be achieved by applying a train of positive pulses (0.65~0.95V/10ns) for potentiation, and continuous negative pulses (-0.8~-1.05V/10ns) for depression (see Supplementary Fig. S1). To avoid the effect of crosstalk in a passive 1R array, we have wired an array of 16 RRAM devices to represent a 4 × 4 RRAM crossbar array in our system demonstration, analogous to a one-transistor one-resistor (1T1R) array. We envision that our proposal can be easily scaled up to larger matrix sizes with a selector attached to each RRAM device in an array.

**Orthogonal Frequency-Division Multiplexing Module**

The RRAM-based DFT module is illustrated in Fig. 2a, 2b, where data are modulated onto non-interfering sub-carriers in the frequency domain. The transformation between the time and frequency domains involves adding/removing a cyclic prefix to avoid inter-sub-carrier interference and IDFT/DFT operations. For the received block of symbols **y**, the DFT of which can be represented as an $N_c$-length vector: $\mathcal{F}(\mathbf{y}) = \mathbf{Wy}$, where $\mathcal{F}(\cdot)$ denotes the DFT operation and the definition of DFT matrix **W** is given in Supplementary Note 5. In the circuit design, the real mapping of DFT matrix, $\mathcal{R}(\mathbf{W})$, is scaled into the RRAM devices' conductance range and stored as the difference between two arrays. The received signal **y** is translated to the input voltages $\mathcal{T}(\mathbf{y})$ for the array. The module computes the DFT of **y**, and the current outputs are the scaled real vector mapping $\mathcal{T}(\mathcal{F}(\mathbf{y}))$. The detailed hardware implementation of this module is provided in Methods. Compared with conventional approaches based on fast Fourier transform (FFT) algorithms[49], the RRAM-based design features the dramatic reduction of computational complexity of from $O(N_c \log N_c)$ for FFT to just a one-step (i.e., $O(1)$) operation. This makes it possible to realize next-generation OFDM communications using extremely large-scale constellations, each comprising thousands to tens-of-thousands of points.

In this section, a simplified single-antenna OFDM system with 4 sub-carriers is demonstrated (see Fig. 2h). The conductance mapping of the DFT matrix to RRAM array is scaled to fit the RRAM devices' conductance range, which are programmed into two arrays. The subtraction of the conductance matrices of these two arrays, in the form of differential pairs, and the corresponding error matrix are presented in Fig. 2c. The block diagrams provided in Fig. 2d-j show the complete signal processing path in the prototypical RRAM-based OFDM system. For the transmitter, a message in bits is firstly modulated into 16-QAM symbols (see Fig. 2h), and then transformed from the frequency domain into the time domain by IDFT (see Fig. 2d, 2e). After adding a cyclic prefix, the OFDM symbols are transmitted over the channel towards the receiver (see Fig. 2e). At the receiver, after removing the cyclic prefix, the RRAM-based DFT is performed to transform the received signal back to the frequency domain (see Fig. 2e, 2f), where the symbols are then demodulated into bits to recover the message. The performance of the receiver with RRAM-based DFT module is experimentally characterized. For a noiseless channel, the constellation diagram at the demapper is shown in Fig. 2i. In this case, the distortion of demodulated symbols is only caused by the inherent noise in RRAM array, with the calculated MER 42dB. For a noisy channel, the constellation diagram at the demapper is shown in Fig. 2j, where the receive signal-noise-ratio (SNR) is 30dB. It is observed that channel noise exacerbates the dispersion of the constellation points by comparison with Fig. 2i. The corresponding MER is calculated as 30dB at which no bit errors occur. More experimental results are provided in Supplementary Fig. S5. Compared with the results from simulation using a double-precision floating-point DFT matrix (see Supplementary Fig. S6 (a)), the performance loss of experimental RRAM-based DFT module due to inherent noise is negligible for a noisy channel. For the noiseless channel, the differences are noticeable while the MER is high enough to ensure errorless demodulation. Hence, in the high SNR regime, the RRAM-based DFT module can reliably recover the message with a zero BER. From Supplementary Fig. S6 (b), one can observe that the curves of software calculation and experimental RRAM-based DFT module are indistinguishable in terms of BER, validating the communication performance of our design.

**Multiple-Input and Multiple-Output Detection Module**

The RRAM-based MIMO detection module is illustrated in Fig. 3a, which spatially multiplexes multiple parallel data-streams. This scales up the system throughput since different symbols are simultaneously transmitted over different antennas. Exploiting the unique channel between each pair of transmit and receive antennas allows each transmitted symbol to be recovered through the module of MIMO detection. In practice, two linear detectors are widely used, namely linear minimum mean square error (L-MMSE) and zero forcing (ZF) detectors (see Supplementary Note 6). They reverse the signal distortion by propagation through a MIMO channel by channel equalization. To be specific, given the channel matrix **H**, the L-MMSE detector minimizes the mean squared error in the estimate of **x** among all linear detectors. The recovered signal vector is given by $\hat{\mathbf{x}} = \left(\mathbf{H}^H\mathbf{H} + \frac{1}{\text{SNR}}\mathbf{I}\right)^{-1}\mathbf{H}^H\mathbf{y}$, where **y** is the received signal vector at the receiver. In hardware implementation, the equivalent real-value channel matrix $\mathcal{R}(\mathbf{H})$ is scaled and written into the RRAM arrays in the way as illustrated in Fig.3a, 3b, and the received signal **y** is scaled and translated to the input voltages $\mathcal{T}(\mathbf{y})$. The output voltages are the real vector mapping $\mathcal{T}(\hat{\mathbf{x}})$, and the details of this circuit are provided in Methods. To cope with heterogeneous propagation environments with different SNRs, the feedback conductance of operational amplifiers can be represented using a RRAM device as shown in Fig. 3b. Our design also applies to ZF detection (see Methods) which solves the least square problem and gives the recovered signal vector as $\hat{\mathbf{x}} = (\mathbf{H}^H\mathbf{H})^{-1}\mathbf{H}^H\mathbf{y}$. As shown in Fig. 3b, the transistor dictates whether L-MMSE or ZF is applied. If the channel matrix is square, i.e., $N_t = N_r = N$, the computational complexity of conventional matrix inversion is $O(N^3)$. The complexity increases rapidly as the number of transmit/receive antennas grows. On the contrary, the proposed MIMO detection performs the computation in just a single step, presenting a promising solution for the 6G ultra-massive MIMO communication with antenna arrays comprising hundreds to thousands of elements.

We experimentally demonstrated the RRAM-based narrowband MIMO system with 2 transmit antennas and 2 receive antennas (see Fig. 3d). The real mapped channel matrix $\mathcal{R}(\mathbf{H})$ is scaled and programmed into the RRAM arrays (see Fig. 3a and Methods). Due to the inherent programming noise of RRAM devices, the actual stored matrix and the corresponding error matrix are shown in Fig. 3c. The experimental results of the constellation diagrams from L-MMSE detection for different transmission powers and a noisy channel (i.e., different SNRs), are shown in Fig. 3e-g. When SNR reduces from 40dB to 30dB to 20dB, the received constellation points become increasingly dispersed from their central points and the corresponding MERs are 33.3dB, 24.5dB and 14.7dB, respectively. More experimental results are provided in Supplementary Fig. S7. Although the MER differences are noticeable in the high SNR regime, such levels are good enough for error-free demodulation. In particular, as shown in Supplementary Fig. S8, the experimental curve using the BER metric is almost indistinguishable from the ideal one. Furthermore, the L-MMSE and ZF methods are compared in Supplementary Fig. S9 with an ill-conditioned MIMO channel, which is emulated by artificially setting the minimum singular value of channel matrix to be much smaller than the other. One can observe that L-MMSE method outperforms ZF in the low SNR regime, but their performance converges at high SNRs.

**Channel Estimation Module**

The RRAM-based channel estimation module is illustrated in Supplementary Fig. S10. To acquire the channel state information (CSI) needed for MIMO detection, the channel matrix is estimated at the receiver using pilot signals that are sent by the transmitter and known a priori to the receiver. Many data symbols can be transmitted between two pilot signals separated by channel coherent time, amortizing the overhead of channel training. A larger ratio between data and pilot symbols improves the system throughput at the cost of adaptively to time-varying channels. The transmitted training matrix $\mathbf{P} \in \mathbb{C}^{N_t \times N_t}$ is known by the receiver, while the actual received matrix is $\mathbf{S} \in \mathbb{C}^{N_r \times N_t}$. By choosing the pilot signal as a unitary matrix[50], i.e., $\mathbf{PP}^H = \mathbf{I}$, the channel matrix estimated by maximum likelihood (ML) or least square (LS) is given as $\hat{\mathbf{H}} = \mathbf{SP}^H$ (see Supplementary Note 7). In the RRAM-based channel estimation module, the real mapped training matrix $\mathcal{R}(\mathbf{P})$ is stored in the RRAM array. Each row vector of the real mapped received matrix $\mathcal{R}(\mathbf{S})$ is translated to the supplied input voltages. The computation can be completed by $2N_r$ read pulses while the complexity is $O(N_r N_t^2)$ for traditional processors. Notably, if we select identity matrix $\mathbf{I}$ as the training matrix, the received matrix is exactly the channel matrix without the need of computation.

When ready, the row vectors of the estimated channel matrix are sequentially written into the RRAM array implementing the MIMO detector. We evaluate the performance of different writing process in terms of system latency. To this end, a mathematical model is developed to facilitate latency analysis for programming an 1T1R array as elaborated in Supplementary Note 8. Consider the writing process using a train of pulses to program an $N \times N$ array in the row-by-row manner. It can be proved that the expected writing latency of write-without-verification and write-with-verification scale with the array size in the way no faster than $O(N\sqrt{\ln N})$ and $O(N \ln N)$, respectively.

**Performance Evaluation of the Complete System**

Recall that we consider the MIMO-OFDM air interface where a transmitter/receiver integrates the RRAM-based OFDM and MIMO modules. In the following, we performed the simulation of a large-scale RRAM-accelerated communication system corresponding to the standard of 5G new radio (NR) (see Supplementary Table 2). To be precise, the flow chart of signal processing of the complete system is illustrated in Supplementary Fig. S13. The simulation of RRAM array programming is based on the RRAM model calibrated using the experimentally acquired device properties such as the evolution of the conductance with voltage pulses (see Supplementary Fig. S1). Since the transmitter is much simpler than the receiver, we focused on the RRAM-based receiver in the remainder of this section. To better demonstrate the performance of our designed in-memory baseband processor, we consider the specific task of transmitting an image chosen from the MNIST dataset as shown in Fig. 4a. Given the SNR being 30dB, the images recovered at the receiver are presented in Fig. 4b and Fig. 4c where RRAM devices are programmed using writing without and with verification schemes, respectively. As a benchmark for comparison, the image resulting from a perfect baseband processor is shown in Fig. 4d. One can observe that the performance of the write-without-verification scheme is poor while the other scheme with verification performs similarly as the ideal processor. To quantify the performance, we present the relation between MER (and BER) and SNR for both schemes as shown in Fig. 4e (and Fig. 4f). The simulation results are aligned with the earlier observation and show that write-with-verification scheme also outperforms the other in terms of communication performance. From the perspective of latency and energy efficiency, where the settings follow Supplementary Table 2 and Supplementary Fig. S1, the performance is compared in Fig. 4g and Fig. 4h. The throughput and energy-efficiency of the proposed in-memory baseband processing exceed those of any reported digital signal processors (DSPs)[51] especially for large antenna arrays. For example, when the MIMO channel is 4 × 4 and the write-with-verification scheme is adopted, the throughput is 32.53TOPS and the energy efficiency is 744.24TOPS/W if channel estimation (and thus the update of MIMO detectors) is performed every sub-frame (1ms). Underpinning the improvements is the ultra-fast one-step baseband processing after channel estimation such that the baseband latency mostly comes from programming the RRAM arrays of MIMO detection module at the beginning of one sub-frame. In contrast, for DSP, data symbols are processed by executing the DFT (or FFT) and MIMO-detection algorithms using digital logic circuits, both suffering from high complexity as discussed (see Supplementary Note 10). Next, there exists a tradeoff between communication performance and latency, i.e., higher performance requires better programming accuracy and thus longer latency. On the one hand, the write-without-verification scheme shows lower latency but poor communication performance in terms of BER and MER, a result of the intrinsic stochasticity of RRAM. On the other hand, RRAM with more states can achieve higher precision but possibly more pulses are needed to reach the target conductance value.

**Discussion**

This work demonstrates the feasibility of UFEE MIMO-OFDM baseband processing by leveraging the emerging in-memory computing technology based on RRAM arrays. The processing latency and energy are mostly contributed by the programming of the RRAM arrays for MIMO detection due to periodic channel estimation, while the following processing of data symbols can be completed in one-step. For a slow fading channel, the infrequent updates of RRAM arrays amplify the advantages of our design in processing time and energy efficiency. In particular, we show that the achievable throughput and energy-efficiency of the RRAM arrays in performing baseband processing can reach 325.27TOPS and 7441.1TOPS/W, respectively, if channel estimation is performed for every 10ms frame. These advantages promise a feasible approach for realizing UFEE baseband processing. It shall be also emphasized that the proposed in-memory baseband processing not only works on RRAM but can be readily applied to other emerging in-memory computing technologies including phase change, ferroelectric and magnetoresistive memories, as detailed in Supplementary Table 3 which lists the device features of our experimental RRAM devices and other types of memristor. There are some observations from the simulation results in Supplementary Fig. S14. First, we compared two different schemes for updating memristor arrays: writing with and without verification, elucidating the importance of verification and low cycle-to-cycle variation in ensuring the accuracy of the operations. To ensure satisfactory communication performance, write-with-verification is suggested for updating the memristor arrays even if the cycle-to-cycle variation is relatively small (e.g., ~0.5%) as shown in the simulation result of programming ferroelectric FET (FeFET) without verification. Second, RRAM can be further improved using ultra-narrow pulse width along with relatively large number of states to achieve ultra-fast conductance updates without compromising the communication performance. For example, the simulation results in Supplementary Fig. S14 show that the UFEE requirements can be met using ferroelectric tunnel junction (FTJ) which is reported for high precision attainable using sub-nanosecond pulses[52]. Leveraging the behavioural model of such memristors, the latency of our in-memory baseband processing system can be reduced to the scale of several microseconds and the energy consumption to the scale of several micro-Jules, which meets the UFEE requirements of 6G communications. Furthermore, in-memory baseband processing is more effective for applications with less stringent precision requirements. For example, if the transmitted messages, such as images, are inputs to the downstream neural networks for inference, the models' robustness against programming noise can ensure high classification accuracy (see Supplementary Note 11). Overall, developing the proposed in-memory baseband processing into a versatile technology is believed to provide a feasible approach for realizing the 6G vision on supporting future services and applications with extremely low latency and high energy-efficiency.

**Methods:**

**RRAM device fabrication:** The RRAM devices have a size of $5 \times 5$ µm$^2$, fabricated on a SiO$_2$ (280nm)-on-Si substrate. After an initial wafer cleaning, the bottom electrode (BE) is formed with lithography followed by Ti (20nm)/Pt (20nm) deposition using ebeam evaporation and a lift-off process. Thereafter, a light inductively coupled plasma (NMC GSE200Plus) etching process in O$_2$ gas atmosphere is used to clean up the photoresist residue on the sidewall of the BE. The HfO$_2$ switching layer is formed using atomic layer deposition (ALD) process (PICOSUN) at 250°C. The deposited HfO$_2$ has a thickness of ~5 nm, as confirmed by both ellipsometer measurements and TEM characterization (see Fig. 1g). The top electrode (TE) is then formed using lithography followed by Ta (20nm)/Pt (20 nm) deposition using ebeam evaporation and a lift-off process. Finally, the contact pads are opened using another step of lithography and wet etch (BOE solution) process.

**Electrical characterization:** The RRAM devices are characterized using a probestation, while the array of RRAM devices is wire-bonded in a ceramic package for system demonstration. Electrical characteristics of the single RRAM are measured using Keysight B1500A semiconductor parameter analyzer. RRAM array programming and readout were achieved using peripheral circuits designed on a custom printed circuit board (PCB); connections between the array of RRAM devices and the PCB were made using wires. We do not envision an issue with the wire parasitic as the array was used only as a proof-of-concept system demonstration and not programmed at high speed. For the operations required by the baseband processor, the parameter analyzer and PCB were connected to two PCs. The auxiliary PC is connected to the parameter analyzer using GPIB cable, and the primary PC (shown in Supplementary Fig. S1) runs C++ and python codes to control the parameter analyzer and the peripheral circuits for programming and readout.

**Read and write operations on RRAM device:** During the write operation, voltage pulses with amplitude of 0.65~0.95V and 10ns were applied to the TE of the RRAM device. The conductance of the RRAM device was modulated upwards using positive voltage pulses. If the conductance of RRAM device is larger than the target, voltage pulses with amplitude of 0.8~1.05V and 10ns were applied using negative pulses. During the read operation, a small read voltage of 0.15V was applied to the RRAM device to avoid cell disturbance. In this work, we used a read after each write operation scheme to ensure the RRAM device's conductance can be set to high accuracy and precision. Multiplication results ($\Sigma V \times G$) were obtained via the summation of the current collected of the desired group of RRAM devices.

**DFT/IDFT and channel estimator circuit:** The DFT/IDFT and channel estimation circuits perform an MVM operation. Consider the circuit with DFT matrix $\mathbf{W} \in \mathbb{C}^{N_c \times N_c}$. The real mapping of the DFT matrix $\mathcal{R}(\mathbf{W}) \in \mathbb{R}^{2N_c \times 2N_c}$ is scaled into the RRAM devices' conductance range by a scaling factor $\alpha$, giving the conductance matrix $\mathbf{G} = \alpha \mathcal{R}(\mathbf{W}) \in \mathbb{R}^{2N_c \times 2N_c}$. The real conductance matrix $\mathbf{G}$ is implemented by the difference between a pair of conductance arrays, $\mathbf{G}^+ - \mathbf{G}^-$, with the utilization of inverting amplifier to invert the voltages. The received signal $\mathbf{y} \in \mathbb{C}^{N_c \times 1}$ is translated to the input voltages with the real vector mapping, such that $\mathbf{v} = \mathcal{T}(\mathbf{y}) \in \mathbb{R}^{2N_c \times 1}$. Leveraging Ohm's law (i.e., current = conductance × voltage), the multiplications $\{G^+_{kl} v_l\}$ and $\{G^-_{kl} v_l\}$ are achieved. Then, Kirchhoff's current law sums these contributions along each row line and the read circuit integrates all the signals, resulting in the current at the $k$-th column $i_k = \sum_{l=1}^{L}(G^+_{kl} - G^-_{kl})v_l$. Therefore, the output currents at the read circuit give the result: $\mathbf{i} = (\mathbf{G}^+ - \mathbf{G}^-)\mathbf{v}$, which gives the DFT result $\alpha \mathcal{T}(\mathbf{x}) = \alpha \mathcal{R}(\mathbf{W})\mathcal{T}(\mathbf{y})$. Since DFT matrix is unitary, i.e., $\mathbf{W}^{-1} = \mathbf{W}^H$, IDFT module circuit is the same as that of DFT when we replace $\mathcal{R}(\mathbf{W})$ with $\mathcal{R}(\mathbf{W})^T$.

**L-MMSE/ZF MIMO detector circuit:** The L-MMSE and ZF MIMO detectors compute the ridge regression and least square estimators, respectively. Consider the L-MMSE detection circuit with channel matrix $\mathbf{H} \in \mathbb{C}^{N_r \times N_t}$. The real mapped channel matrix $\mathcal{R}(\mathbf{H}) \in \mathbb{R}^{2N_r \times 2N_t}$ is scaled into the RRAM devices' conductance range by a scaling factor $\alpha$, giving the conductance matrix $\mathbf{G} = \mathbf{G}^+ - \mathbf{G}^- = \alpha \mathcal{R}(\mathbf{H}) \in \mathbb{R}^{2N_r \times 2N_t}$ which is implemented as the difference between two RRAM arrays. The real vector mapping of the received signal $\mathcal{T}(\mathbf{y}) \in \mathbb{R}^{2N_r \times 1}$ is translated to input currents. To make the voltages in the circuit within a reasonable range, the input currents are also scaled as $\mathbf{i} = \alpha \mathcal{T}(\mathbf{y}) \in \mathbb{R}^{2N_r \times 1}$. The two arrays at the left-hand side constitute the conductance matrix $-\mathbf{G} = \mathbf{G}^- - \mathbf{G}^+$ with voltages $\mathbf{v}$ supplied at the bottom of the nether array. The Kirchhoff's current law sums the output currents from the left RRAM array pair, $-\mathbf{Gv}$, and the input currents, $\mathbf{i}$, such that the input currents at the operational amplifiers are $\mathbf{i}' = -\mathbf{Gv} + \mathbf{i}$. Hence, the output voltages that supplied to the right RRAM array pair are $\mathbf{v}' = -\frac{\mathbf{i}'}{g_1} = \frac{\mathbf{Gv}-\mathbf{i}}{g_1}$, where $g_1$ is the feedback conductance of the TIAs. Then, the right RRAM array pair, whose conductance matrix is represented by $\mathbf{G}^T = (\mathbf{G}^+ - \mathbf{G}^-)^T$, performs the MVM computation and outputs the current vector $\mathbf{i}'' = \mathbf{G}^T\mathbf{v}' = \mathbf{G}^T\frac{\mathbf{Gv}-\mathbf{i}}{g_1}$. The currents are applied to the other set of TIAs, so that $\mathbf{i}'' = -g_2\mathbf{v}$, where $g_2$ is the feedback conductance of the TIAs in this set. Accordingly, one can observe the relation: $\mathbf{G}^T\frac{\mathbf{Gv}-\mathbf{i}}{g_1} = -g_2\mathbf{v}$, which gives the output voltages $\mathbf{v} = (\mathbf{G}^T\mathbf{G} + g_1g_2\mathbf{I})^{-1}\mathbf{G}^T\mathbf{i}$. By setting the SNR as $\alpha^2(g_1g_2)^{-1}$, the designed L-MMSE circuit outputs the desired vector: $\mathcal{T}(\hat{\mathbf{x}}) = \left(\mathcal{R}(\mathbf{H})^T\mathcal{R}(\mathbf{H}) + \frac{1}{\text{SNR}}\mathbf{I}_{2N_t\times 2N_t}\right)^{-1}\mathcal{R}(\mathbf{H})^T\mathcal{T}(\mathbf{y})$. When the feedbacks of the TIAs in the second set are open, i.e., $g_2 = 0$, the output voltages of the circuit are $\mathbf{v} = (\mathbf{G}^T\mathbf{G})^{-1}\mathbf{G}^T\mathbf{i}$. This computes the ZF and gives the desired vector $\mathcal{T}(\hat{\mathbf{x}}) = \left(\mathcal{R}(\mathbf{H})^T\mathcal{R}(\mathbf{H})\right)^{-1}\mathcal{R}(\mathbf{H})^T\mathcal{T}(\mathbf{y})$.

**Data availability:**

All data supporting the findings of this study are available from the corresponding author on request.

**Additional information:**

Supplementary Information accompanies this paper.

**Competing interests**:

The authors declare no competing financial interests.

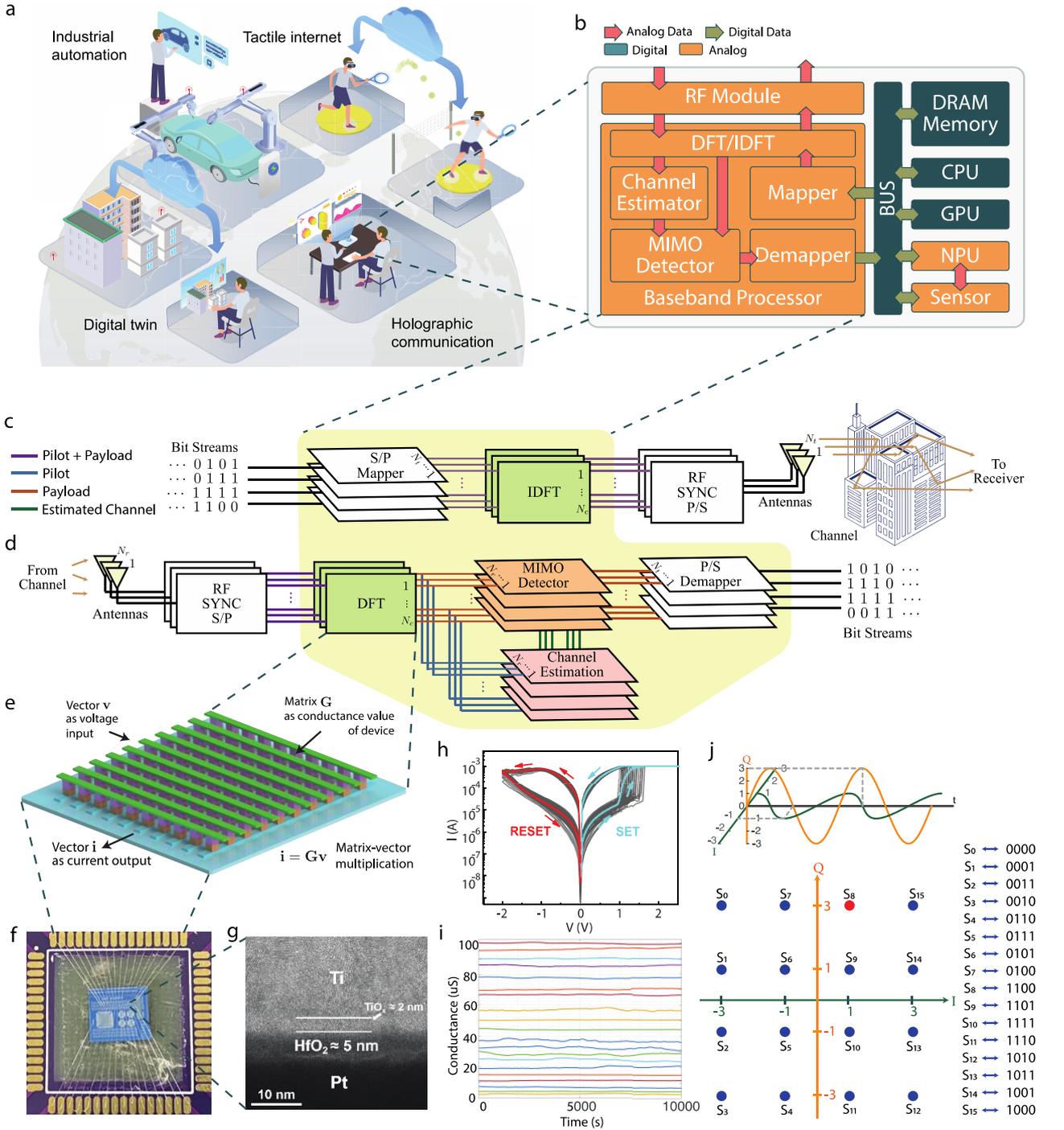

**Fig. 1: Next generation wireless communications and baseband processors.** (**a**) Some key use cases for 6G communication including industrial automation, tactile internet, holographic communications, and digital twin. (**b**) Proposed next-generation system-on-chip (SoC) in a hybrid analogue-and-digital architecture. The basic components include CPU, GPU, neural processing unit (NPU), sensor, baseband processor, and radio modems, where both baseband processing and NPU can be accelerated by resistive Random-Access Memory (RRAM) arrays. The illustrated baseband processor aims at the Multiple-Input-Multiple-Output Orthogonal Frequency-Division Multiplexing (MIMO-OFDM) air interface for multi-antenna broadband wireless communications. (**c**) The architecture of RRAM-based transmitter. It consists of baseband processing modules [i.e., mapper and Inverse Discrete Fourier Transform (IDFT)], Radio Frequency (RF) modem, and an array of transmit antennas. Each layer represents a piece of RRAM-based circuit. (**d**) The architecture of RRAM-based receiver. It is comprised of an array of receive antennas, RF front-end, and baseband processing modules [i.e., Discrete Fourier Transform (DFT), channel estimation, MIMO detection, and demapper]. (**e**) An illustration of the RRAM array. It is the basic component in the design. (**f**) A photo of the RRAM array. (**g**) The zoomed-in Transmission Electron Microscopy (TEM) image of the RRAM device, with material stack as indicated. (**h**) RRAM current-voltage (I-V) characteristics. (**i**) Room-temperature state retention and read disturb of the device states. (**j**) The constellation diagram of 16 quadrature amplitude modulation (16-QAM). The bit stream is split into in-phase (denoted by $I$) and quadrature (denoted by $Q$) streams. $I$ and $Q$ components are Gray encoded, where neighbour points only differ in 1-bit position, to produce symbol points in the constellation.

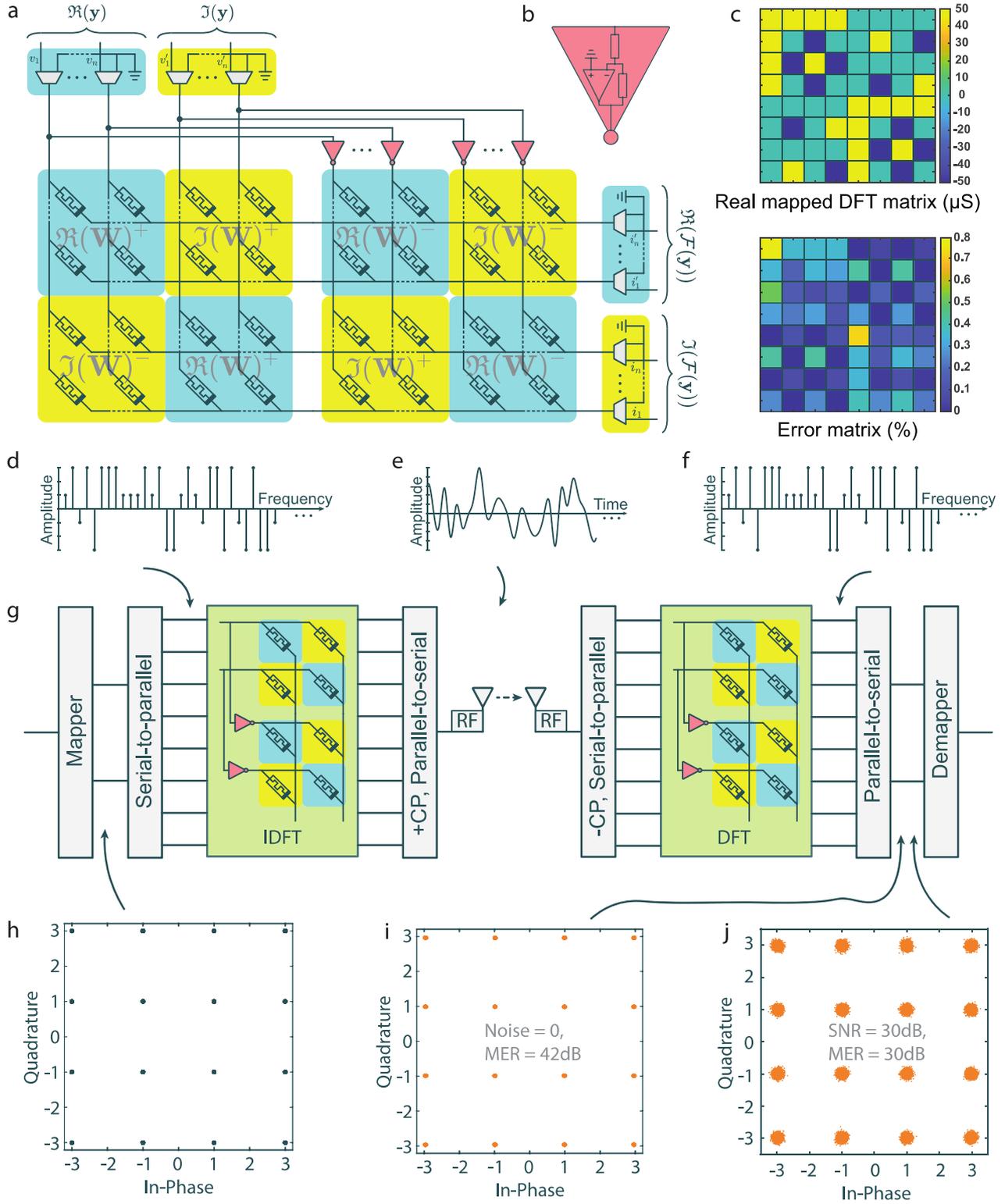

**Fig. 2: Orthogonal frequency-division multiplexing modules.** (a) The architecture of RRAM-based DFT module. The DFT matrix **W** is stored in the RRAM array and the input signal **y** is translated as the voltages to be applied to the array. The elements (and signals) in real and imaginary domains are highlighted by different colours. The module performs the DFT over signal **y** and the read drivers get the result $\mathcal{F}(\mathbf{y}) = \mathbf{W}\mathbf{y}$. (b) Inverting amplifier. (c) In the experiment, the real mapping of the DFT matrix is scaled and programmed in RRAM array: (upper) conductance matrix and (lower) corresponding error matrix. (d)-(j) A single-antenna OFDM system in the demonstration. At the transmitter, (h) the message is firstly modulated from bits to symbols by 16-QAM, and then (d) transformed from frequency domain to time domain by IDFT module. (e) The OFDM symbols are transmitted over the channel to the receiver. At the receiver, (f) DFT is performed to transform the received signal from time domain to frequency domain. (i) The symbols are demodulated into bits and the constellation diagram is recovered without channel noise. (j) The symbols are demodulated into bits and the constellation diagram is recovered with channel noise.

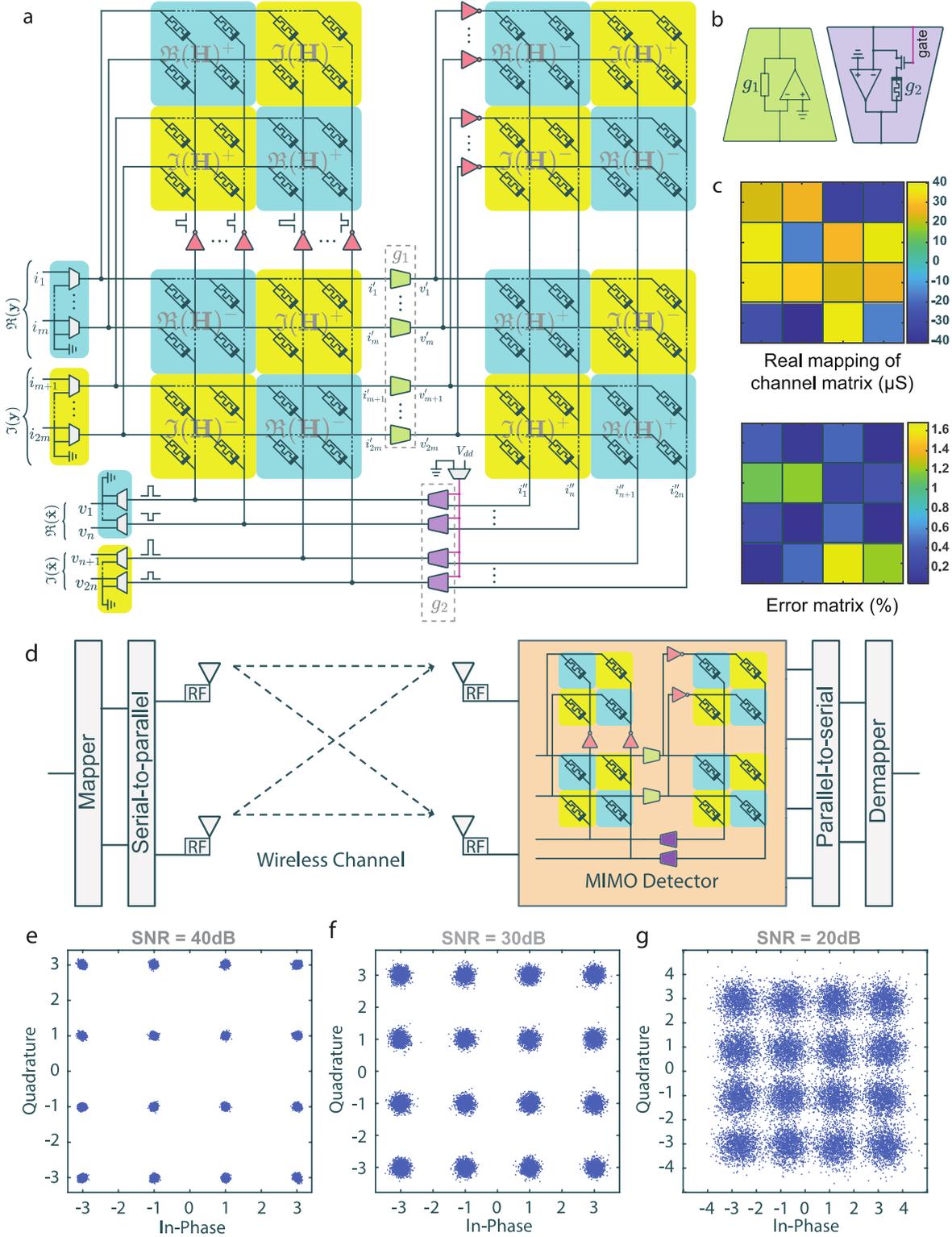

**Fig. 3: Multiple-input and multiple-output modules. (a)** The architecture of RRAM-based MIMO detection module. The channel matrix **H** is stored in the four RRAM arrays as marked in the figure and the input signal **y** is scaled and translated as the input currents. The elements (and signals) in real and imaginary domains are highlighted by different colours. For linear minimum mean square error (L-MMSE) detector, the circuit gives the recovered signal vector $\hat{\mathbf{x}} = \left(\mathbf{H}^H\mathbf{H} + \frac{1}{\text{SNR}}\right)^{-1}\mathbf{H}^H\mathbf{y}$. For zero forcing (ZF) detector, it gives $\hat{\mathbf{x}} = (\mathbf{H}^H\mathbf{H})^{-1}\mathbf{H}^H\mathbf{y}$. **(b)** The transistor controls whether L-MMSE or ZF modules is adopted. When the gate is grounded, the circuit performs ZF detection. Otherwise, L-MMSE is selected. In addition, to adapt to environments with different SNRs, the feedback conductance of the operational amplifiers is tuneable as shown. **(c)** In the experiment, the real mapping of the channel matrix is scaled and programmed in RRAM array: (upper) conductance matrix and (lower) corresponding error matrix. **(d)** A single-carrier MIMO system in the demonstration. **(e)-(g)** The recovered symbols from RRAM-based L-MMSE MIMO detector in the experiment when SNR is 40dB, 30dB and 20dB, respectively.

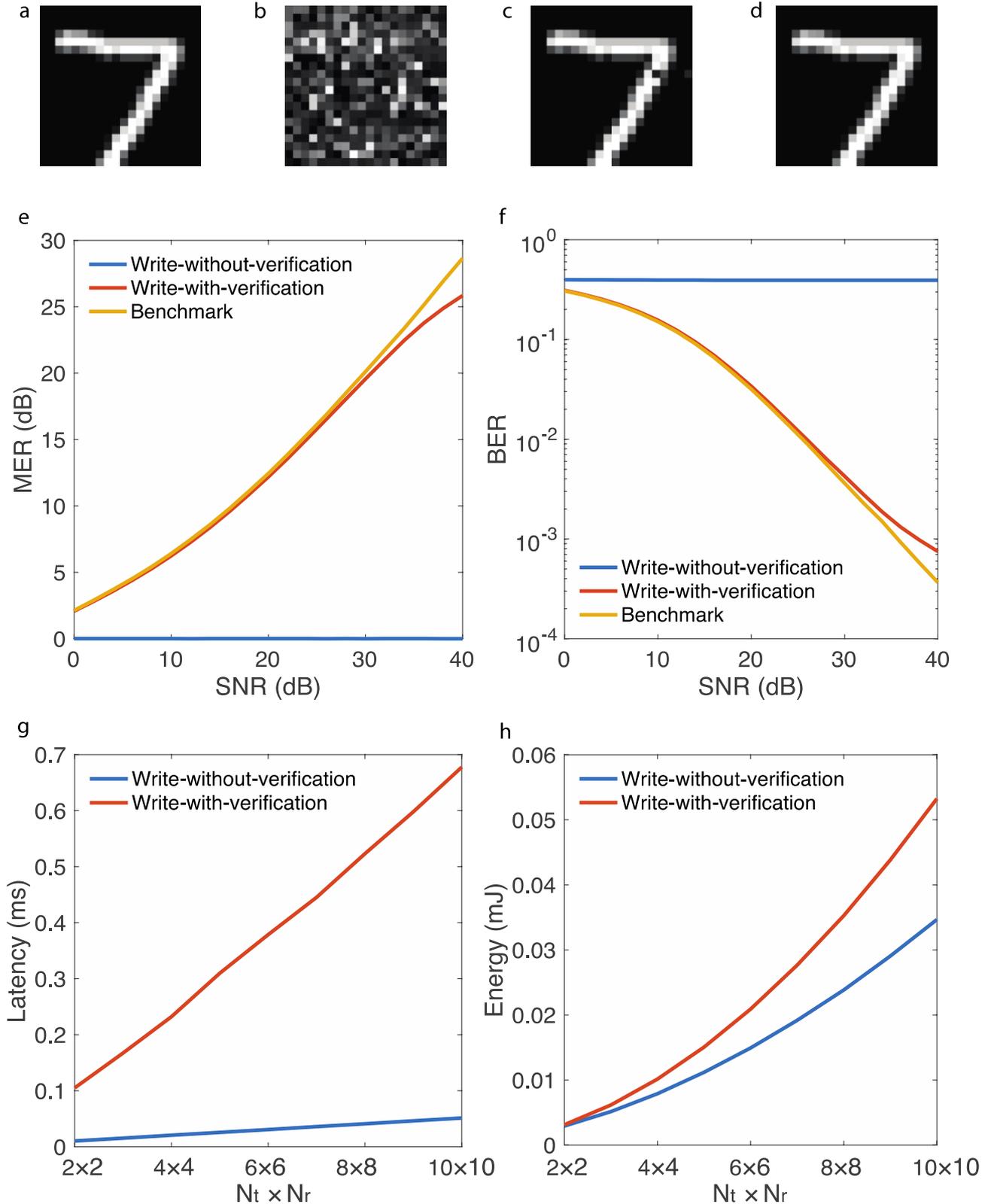

**Fig. 4: Performance evaluation of in memory MIMO-OFDM baseband processing using experimental RRAMs.** The simulations target a large-scale MIMO-OFDM system of 1024 sub-carriers, 4 transmit antennas and 4 receive antennas unless specified otherwise. The behavioural model of memristor is based on the experimental testing of HfOx-based RRAM devices. **(a-d)** An illustration of the communication performance of transmitting an image over a noisy channel (SNR = 30dB). **(a)** The original image. **(b, c)** The recovered images are from RRAM-based baseband processor where the RRAM arrays at the MIMO detector are programmed by **(b)** write without verification and **(c)** write with verification schemes. **(d)** Benchmark: software result. **(e, f)** Under different channel conditions, the resultant **(e)** Modulation Error Ratio (MER) and **(f)** Bit Error Ratio (BER) from the three schemes. **(g, h)** For write with and without verification schemes, the **(g)** latency and **(h)** energy are evaluated in terms of different MIMO sizes.